\documentclass[onecolumn]{revtex4}
\usepackage{ragged2e}
\usepackage{amsmath}
\usepackage{graphicx}
\usepackage{bm}

\usepackage[usenames,dvipsnames]{color}

\definecolor{darkblue}{RGB}{0,0,196}
\usepackage[colorlinks=true,linkcolor=darkblue,citecolor=darkblue,urlcolor=darkblue]{hyperref}

\begin{document}

\title{Comparing effective temperatures in standard and Tsallis distributions
from transverse momentum spectra in small collision systems}
\vspace{0.5cm}

\author{Peng-Cheng Zhang$^{1,}$\footnote{202312602003@email.sxu.edu.cn},
Pei-Pin Yang$^{2,}$\footnote{peipinyangshanxi@163.com;
peipinyang@xztu.edu.cn}, Ting-Ting
Duan$^{1,}$\footnote{202312602001@email.sxu.edu.cn}, Hailong
Zhu$^{1,}$\footnote{Correspondence: zhuhl@sxu.edu.cn}, Fu-Hu
Liu$^{1,}$\footnote{Correspondence: fuhuliu@163.com;
fuhuliu@sxu.edu.cn}, Khusniddin K.
Olimov$^{3,4,}$\footnote{Correspondence: khkolimov@gmail.com;
kh.olimov@uzsci.net}}

\affiliation{\it{$^1$Institute of Theoretical Physics \& College
of Physics and Electronic Engineering, \\ State Key
Laboratory of Quantum Optics Technologies and Devices \& \\
Collaborative Innovation Center of Extreme Optics, Shanxi
University, Taiyuan 030006, China
\\
$^2$Department of Physics, Xinzhou Normal University, Xinzhou
034000, China
\\
$^3$Laboratory of High Energy Physics, Physical-Technical
Institute of Uzbekistan Academy of Sciences, Chingiz Aytmatov Str.
2b, Tashkent 100084, Uzbekistan
\\
$^4$Department of Natural Sciences, National University of Science
and Technology MISIS (NUST MISIS), Almalyk Branch, Almalyk 110105,
Uzbekistan}}

\begin{abstract}

\vspace{0.5cm}

\noindent {\bf Abstract:} The transverse momentum ($p_T$) spectra
of identified light charged hadrons, specifically bosons
($\pi^{\pm}$ and $K^{\pm}$) as well as fermions [$p(\bar p)$],
produced in small collision systems, namely deuteron-gold (d+Au)
and proton-proton (p+p) collisions at the top energy of the
Relativistic Heavy Ion Collider (RHIC) with a center-of-mass
energy of $\sqrt{s_{NN}}=200$ GeV, are investigated in this paper.
In present study, d+Au collisions are categorized into three
centrality classes: central (0--20\%), semi-central (20--40\%),
and peripheral (40--100\%) collisions. Various types of
distributions, including standard [Bose-Einstein (Fermi-Dirac) and
Boltzmann] and Tsallis distributions, are employed to fit the same
$p_T$ spectra to derive different effective temperatures denoted
as $T_{eff}$. The results indicate that $T_{eff}$ values obtained
from Bose-Einstein, Boltzmann, Fermi-Dirac, and Tsallis
distributions exhibit systematically a decreasing trend.
Meanwhile, these $T_{eff}$ values also show a decreasing trend
with a decrease in collision centrality. Furthermore, based on the
spectra of given particles, a perfect linear relationship is
observed between different pairwise combinations of $T_{eff}$
derived from both Boltzmann and Bose-Einstein (Fermi-Dirac)
distributions as well as between Tsallis and Bose-Einstein
(Fermi-Dirac) distributions.
\\
\end{abstract}

\maketitle

\parindent=15pt

\section{Introduction}

In high-energy collisions, temperature serves as a crucial
concept~\cite{1}. It is essential not only for understanding
collision mechanisms and particle
production~\cite{1a,1b,1c,1d,1e}, but also for investigating the
critical point associated with phase transitions from hadronic
matter to quark-gluon plasma (QGP or quark matter), followed by
quick hadronization back to hadronic matter~\cite{2,3,4,5,6}.
Temperature plays an irreplaceable role in these
processes~\cite{7,8,9,10,10a}. It is widely accepted that if a
first-order phase transition occurs with increasing collision
energy, the excitation function of temperature will display either
a plateau or saturated structure~\cite{10b}. Conversely, if the
phase transition is characterized by crossover behavior (where
changes occur gradually), the excitation function may show a slow
upward trend.

There are various types of temperatures associated with
high-energy collisions, including, but not limited to, initial
temperature, chemical freeze-out temperature, kinetic freeze-out
temperature, and effective temperature~\cite{11,12,13,14,15}. The
initial temperature characterizes the level of excitation within
the collision system during its early stage when gluons exist in a
saturated state. The chemical (kinetic) freeze-out temperature
reflects the degree of excitation during the chemical (kinetic)
freeze-out phase when particle ratios (momentum distributions)
remain fixed. The effective temperature ($T_{eff}$ or simply $T$)
serves as an overall indicator of thermal motion among particles
in all directions, and collective motion within the collision
system along the radial direction, at the kinetic freeze-out
stage. In this context, thermal motion can be described by the
kinetic freeze-out temperature while collective motion is
represented by either the average radial flow velocity or its
transverse component: the average transverse flow velocity of
particles.

It is important to note that these temperatures are
model-dependent. A standard baseline is essential for comparing
similar types of temperatures derived from different models or
distributions. Indeed, distinct types of temperatures necessitate
different standard baselines. Investigating these standard
baselines for various types of temperatures represents a
substantial and challenging endeavor that cannot be adequately
addressed within a single paper. Instead, as foundational work in
this area, we aim to compare $T$ values obtained from different
models or distributions. Both Bose-Einstein and Fermi-Dirac
distributions stem from widely utilized ideal gas model and
(Maxwell-)Boltzmann distribution provides their approximate
representation. Furthermore considering that Tsallis
distribution~\cite{16,17,18,19,20} effectively constitutes an
infinite multi-component Boltzmann distribution with appropriate
weights, these distributions have been selected as illustrative
examples to compare their $T$ for our study.

The extraction of different $T$ from the same transverse momentum
($p_T$) spectra of identified hadrons in the final state of
high-energy collisions originates from different distribution
functions. While numerous distributions can be employed to derive
a series of $T$, these values differ due to their model
dependence~\cite{21,22}. In this paper, we utilize standard
[Bose-Einstein (Fermi-Dirac) and Boltzmann] and Tsallis
distributions to extract distinct $T$ values from identical
experimental $p_T$ spectra. We also establish the relationships
between different pairwise $T$.

The structure of this paper is organized as follows: Section 2
describes the formalism associated with various types of
distributions. Results and discussions are presented in Section 3.
Finally, we provide our summary and conclusions in Section 4.

\section{Different types of distributions}

For a given emission source of particles produced in high-energy
collisions, one may apply different models or distributions to
characterize the distribution laws for these particles. The
relativistic ideal gas model, widely utilized in thermodynamics,
statistical physics, and quantum mechanics, is favored in this
study. The invariant yield or particle momentum ($p$) distribution
is expressed as~\cite{23}
\begin{align}
E\frac{d^3N}{d^3p}&= \frac{1}{2\pi p_T}\frac{d^2N}{dydp_T}
= \frac{1}{2\pi m_T}\frac{d^2N}{dydm_T} \nonumber\\
&= \frac{gV}{(2\pi)^3}E
\bigg[\exp\bigg(\frac{E-\mu}{T}\bigg)\mp1\bigg]^{-1}.
\end{align}
Here, $N$ represents the particle number, $g$ denotes the
degeneracy factor, $\mu$ indicates the chemical potential, which
is close to zero at high energy, $E=\sqrt{p^2+m_0^2}=m_T\cosh y$
refers to energy, $m_T=\sqrt{p_T^2+m_0^2}$ is transverse mass,
$m_0$ signifies rest mass, $y=(1/2)\ln[(E+p_z)/(E-p_z)]$ describes
rapidity, $p_z$ shows the longitudinal momentum of considered
particle, and $V$ represents the volume of a collision system.
Additionally, $-1$ in Eq. (1) corresponds to Bose-Einstein
distribution applicable for bosons such as $\pi^{\pm}$ and
$K{^\pm}$ (with $g=1$), while $+1$ in Eq. (1) pertains to
Fermi-Dirac distribution relevant for fermions like $p(\bar p)$
(with $g=2$).

The density function of momenta $p$ can be given by expression
\begin{align}
\frac{dN}{dp}=\frac{2gV}{(2\pi)^2} p^2
\bigg[\exp\bigg(\frac{\sqrt{p^2+m_0^2}-\mu}{T}\bigg)\mp1\bigg]^{-1}.
\end{align}
The unit-density function of $y$ and $p_T$ is written as~\cite{23}
\begin{align}
\frac{d^2N}{dydp_T} =&\, \frac{gV}{(2\pi)^2}p_T
\sqrt{p_T^2+m_0^2}\cosh y \nonumber\\
&\times \bigg[\exp\bigg(\frac{\sqrt{p_T^2+m_0^2}\cosh
y-\mu}{T}\bigg)\mp1\bigg]^{-1}.
\end{align}
The density function of $p_T$ is
\begin{align}
\frac{dN}{dp_T} =&\,\frac{gV}{(2\pi)^2} p_T \sqrt{p_T^2+m_0^2}
\int_{y_{\min}}^{y_{\max}} \cosh y \nonumber\\
&\times \bigg[\exp\bigg(\frac{\sqrt{p_T^2+m_0^2}\cosh
y-\mu}{T}\bigg)\mp1\bigg]^{-1}dy,
\end{align}
where $y_{\min}$ and $y_{\max}$ denote the minimum and maximum
rapidities, respectively, in the rapidity bin
$[y_{\min},y_{\max}]$ measured in experiments.

Generally, the experimental range $[y_{\min},y_{\max}]$
encompasses mid-rapidity ($y_C=0$), and the difference $\Delta
y=y_{\max}-y_{\min}$ is not excessively large ($\Delta y<1\sim2$).
If the interval $[y_{\min},y_{\max}]$ does not include
mid-rapidity, it can be shifted to center around mid-rapidity by
adding or subtracting a certain quantity. Specifically,
$[y_{\min},y_{\max}]$ can be directly adjusted to $[-\Delta
y/2,\Delta y/2]$, allowing for the subtraction of kinetic energy
associated with directional motion along the longitudinal axis
from the total energy of particles. While radial flow effects do
influence $T$, contributions from directional motion can be
effectively eliminated through shift of rapidity.

In relation to the aforementioned Bose-Einstein (Fermi-Dirac)
distribution expressed in terms of invariant yield, $p$ ($p_T$)
density, and unit-density concerning $y$ and $p_T$, removing
factors of $\mp1$ from these distributions yields approximate
expressions suitable for indistinguishable bosons and fermions,
namely leading to Boltzmann distribution form. For classical
particle emission sources, employing a Boltzmann distribution can
be appropriate.

Given the complexity inherent in high-energy collision processes,
multiple emission sources can exist for produced particles.
Consequently, one might consider a multi-source thermal
model~\cite{24,25,26,27,28}, wherein different sources operate at
varying excitation levels or engage distinct reaction mechanisms.
The previously discussed Bose-Einstein (Fermi-Dirac) or Boltzmann
distributions thus transform into a multi-component distribution
that reflects a multi-regional structure within $p_T$
spectra~\cite{29,30,31}.

Due to commonalities and similarities~\cite{32,33,34,35}, as well
as universality~\cite{36,37,38,39} observed in high-energy
collisions, various components within this multi-component
distribution can adopt analogous forms. Regarding the probability
density function of $p_T$ as an example, the multi-component
distribution is expressed as
\begin{align}
\frac{1}{N}\frac{dN}{dp_T} =\sum_{i=1}^{n_0}
k_i\frac{1}{N_i}\frac{dN_i}{dp_T},
\end{align}
where $i$ denotes the $i$-th component in $n_0$ components, $N_i$
is the number of particles in the $i$-th component, and $k_i$ is
the fraction of the $i$-th component. Naturally, the normalization
$\sum_{i=1}^{n_0}k_i=1$ is obeyed.

The multi-component distribution shows a multi-temperature, i.e.,
a temperature fluctuation. The average temperature from the
multi-component distribution can be calculated as
\begin{align}
T=\sum_{i=1}^{n_0}k_iT_i,
\end{align}
where $T_i$ is the temperature obtained from the $i$-th component.
It is worth emphasizing that each component (the $i$-th component)
in the multi-component $p_T$ distribution should be the
probability density function [$(1/N_i)dN_i/dp_T$] which is
normalized to 1, but not the general density function
($dN_i/dp_T$), which is normalized to $N_i$.

If one considers other types of distributions, but not the
probability density function, the multi-component distributions
are represented as
\begin{align}
E\frac{d^3N}{d^3p}=\sum_{i=1}^{n_0} E\frac{d^3N_i}{d^3p},
\end{align}
\begin{align}
\frac{dN}{dp}=\sum_{i=1}^{n_0} \frac{dN_i}{dp},
\end{align}
\begin{align}
\frac{d^2N}{dydp_T}=\sum_{i=1}^{n_0} \frac{d^2N_i}{dydp_T},
\end{align}
\begin{align}
\frac{dN}{dp_T}=\sum_{i=1}^{n_0} \frac{dN_i}{dp_T}.
\end{align}
The ratio of the contribution of the $i$-th component to that of
$n_0$ components is exactly the fraction of the $i$-th component,
$k_i$, which does not appear obviously in these multi-component
distributions.

In the case of $n_0$ being infinitely large, the multi-component
Boltzmann distribution results in the Tsallis distribution which
has the following forms~\cite{16,17,18,19,20}
\begin{align}
E\frac{d^3N}{d^3p} &= \frac{1}{2\pi p_T}\frac{d^2N}{dydp_T}
= \frac{1}{2\pi m_T}\frac{d^2N}{dydm_T} \nonumber\\
&=\frac{gV}{(2\pi)^3}E
\bigg[1+(q-1)\frac{E-\mu}{T}\bigg]^{-\frac{q}{q-1}},
\end{align}
\begin{align}
\frac{dN}{dp}=\frac{2gV}{(2\pi)^2} p^2
\bigg[1+(q-1)\frac{\sqrt{p^2+m_0^2}-\mu}{T}\bigg]^{-\frac{q}{q-1}},
\end{align}
\begin{align}
\frac{d^2N}{dydp_T} =&\, \frac{gV}{(2\pi)^2}p_T
\sqrt{p_T^2+m_0^2}\cosh y \nonumber\\
&\times \bigg[1+(q-1)\frac{\sqrt{p_T^2+m_0^2}\cosh y-\mu}{T}
\bigg]^{-\frac{q}{q-1}},
\end{align}
\begin{align}
\frac{dN}{dp_T} =&\, \frac{gV}{(2\pi)^2} p_T \sqrt{p_T^2+m_0^2}
\int_{y_{\min}}^{y_{\max}} \cosh y \nonumber\\
&\times \bigg[1+(q-1)\frac{\sqrt{p_T^2+m_0^2}\cosh y-\mu}{T}
\bigg]^{-\frac{q}{q-1}}dy.
\end{align}
Here, $q$ is the entropy index, which describes the degree of
non-equilibrium of the collision system.

We initially attempted to fit experimental data using a
single-component standard distribution, but encountered
unsatisfactory results. Subsequently, we employed a two-component
standard distribution for fitting; however, the outcomes remained
not satisfactory. When utilizing a three-component standard
distribution, the results became acceptable. Consequently, it was
necessary to adopt the three-component standard distribution.
Nevertheless, with the introduction of the entropy index $q$, a
single Tsallis distribution effectively smoothed temperature
fluctuations inherent in the three-component standard distribution
and achieved satisfactory fitting of the experimental data.
Therefore, a single-component Tsallis distribution suffices,
leaving two- or three-component Tsallis distributions unnecessary.

In the present work, the probability density function
$[(1/N)(dN/dp_T)]$ is used first of all in the calculation of
multi-component standard distribution. To fit the experimental
invariant yield, $(1/2\pi p_T) d^2N/dydp_T$, one needs to use the
relation
\begin{align}
\frac{1}{2\pi p_T}\frac{N_0}{N}\frac{dN}{dp_T}=\frac{1}{2\pi p_T}
\int_{y_{\min}}^{y_{\max}} \frac{d^2N}{dydp_T}dy,
\end{align}
where $N_0$ is the normalization constant that is generally the
area under the data, $dN/dp_T$. In the fit, $N_0$ is determined by
the data itself and has no relation to the model.

For the single-component Tsallis distribution, the unit density
function, $d^2N/dydp_T$, of $y$ and $p_T$ can be utilized. To fit
the experimental invariant yield, given by $(1/2\pi p_T)
d^2N/dydp_T$, one can conveniently multiply $d^2N/dydp_T$ in Eq.
(13) by $(1/2\pi p_T)$. This is justified because for small values
of rapidity, specifically when $y\approx0$, it follows that $\cosh
y\approx1$, due to the narrow rapidity bin defined as
$[y_{\min},y_{\max}]$, which encompasses mid-rapidity.

In some cases, to differentiate between various effective
temperatures, $T_{\text{Bose-Einstein}}$
($T_{\text{Fermi-Dirac}}$), $T_{\text{Boltzmann}}$, and
$T_{\text{Tsallis}}$ are used to denote those obtained from their
respective distributions: Bose-Einstein (Fermi-Dirac), Boltzmann,
and Tsallis distributions. Regardless of the symbol employed in
our calculations, while acknowledging that contributions from
collective motion within the collision system or flow effects of
produced particles may influence the temperature parameter, we aim
to derive an effective temperature corresponding to the kinetic
freeze-out during evolution of the system.

\section{Results and discussion}

\subsection{Comparison with experimental data}

Figures 1--3 present the invariant yields, $(1/2\pi
p_T)d^2N/dydp_T$, of (a) positive and negative pions ($\pi^+$ and
$\pi^-$), (b) positive and negative kaons ($K^+$ and $K^-$), as
well as (c) protons and antiprotons ($p$ and $\bar{p}$), produced
in mid-rapidity range ($|y|<0.5$) in deuteron-gold (d+Au)
collisions across three centrality percentage classes (central
0--20\%, semi-central 20--40\%, and peripheral 40--100\%) and in
proton-proton (p+p) collisions at a center-of-mass energy of
$\sqrt{s_{NN}}=200$ GeV, which is the top energy per nucleon pair
at the Relativistic Heavy Ion Collider (RHIC). The closed and open
symbols represent experimental data collected by the STAR
Collaboration~\cite{40}, while the solid and dotted curves
correspond to our results for positive and negative hadrons fitted
using (i) three-component Bose-Einstein and Boltzmann
distributions for $\pi^++\pi^-$ ($K^++K^-$), (ii) three-component
Fermi-Dirac and Boltzmann distributions for $p+\bar p$, and (iii)
single-component Tsallis distribution for all hadrons mentioned
above.

In Figures 1--3, in the fit of three-component Bose-Einstein and
Boltzmann distributions for the spectra of $\pi^++\pi^-$
($K^++K^-$), as well as in the fit of three-component Fermi-Dirac
and Boltzmann distributions for the spectra of $p+\bar p$, the
normalization constant is $N_0$ in Eq. (15). At the same time, the
single-component Tsallis distribution is applied for the mentioned
spectra, where the normalization constant is $V$ in Eq. (13).
Although these two normalization constants could be unified
theoretically, they arise from different calculation
methodologies. Notably, both $N_0$ and $V$ are intrinsically
determined by the data itself. It should be noted that $N_0$ is
also applicable to the single-component Tsallis distribution.
Similarly, $V$ can be utilized for both three-component
Bose-Einstein (Fermi-Dirac) distributions as well as Boltzmann
distributions due to their reliance on identical datasets, albeit
minor deviations may occur regarding normalizations.

\begin{figure*}[htb!]
\begin{center}
\includegraphics[width=15.5cm]{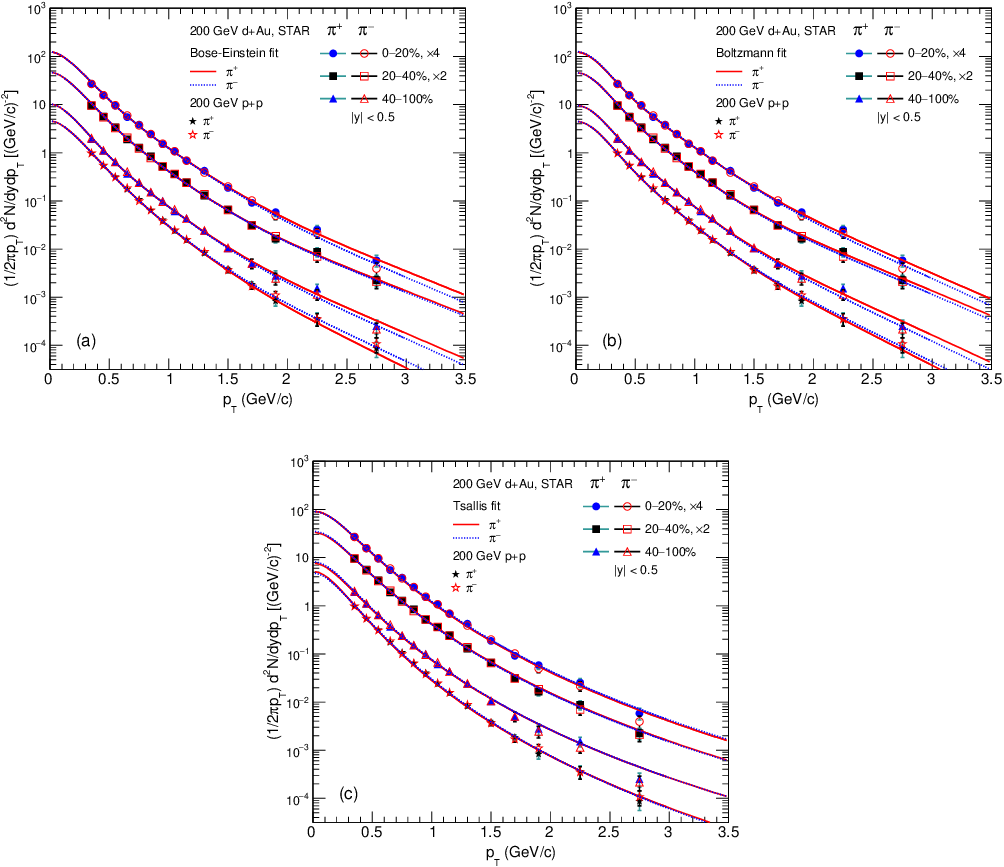}
\end{center}
\justifying\noindent {Figure 1. The invariant yields, $(1/2\pi
p_T)d^2N/dydp_T$, of $\pi^+$ and $\pi^-$ produced in $|y|<0.5$ in
d+Au collisions with three centrality percentages and in p+p
collisions at $\sqrt{s_{NN}}=200$ GeV. The closed and open symbols
represent the experimental data measured by the STAR
Collaboration~\cite{40}. The solid and dotted curves are our
results for $\pi^+$ and $\pi^-$ fitted by the (a) three-component
Bose-Einstein, (b) three-component Boltzmann, and (c)
single-component Tsallis distributions, respectively.}
\end{figure*}

\begin{figure*}[htb!]
\begin{center}
\includegraphics[width=15.5cm]{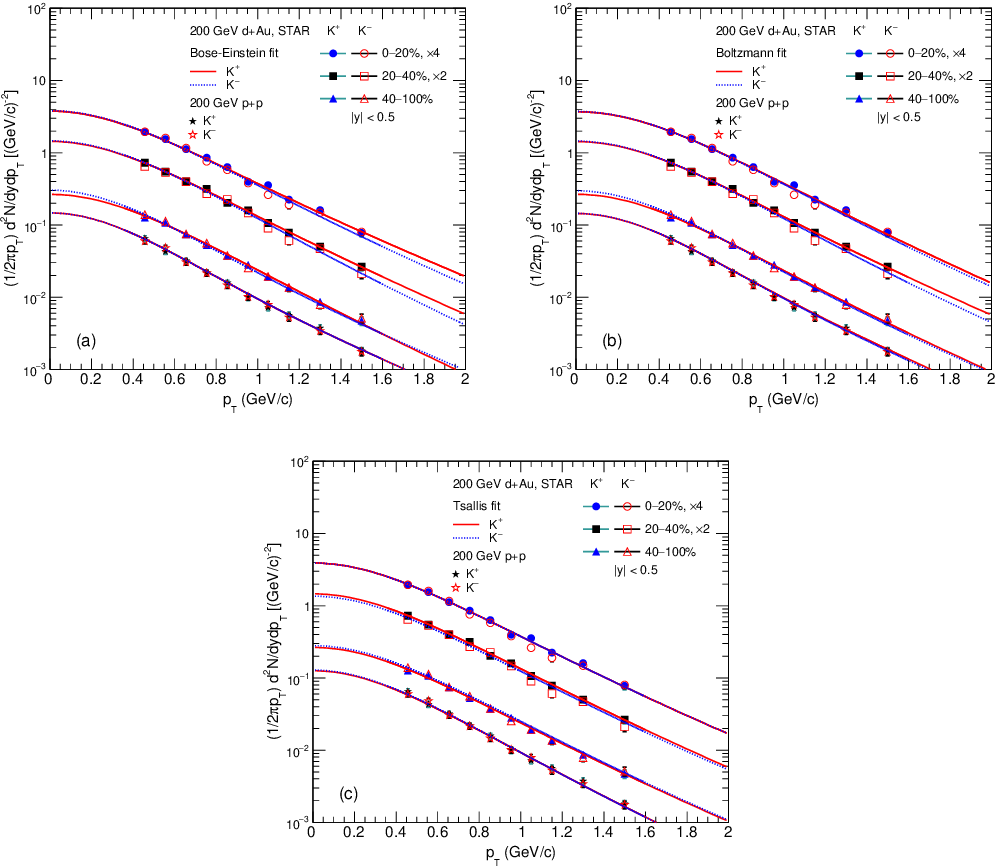}
\end{center}
\justifying\noindent {Figure 2. The invariant yields, $(1/2\pi
p_T)d^2N/dydp_T$, of $K^+$ and $K^-$ produced in $|y|<0.5$ in d+Au
collisions with three centrality percentages and in p+p collisions
at $\sqrt{s_{NN}}=200$ GeV. The closed and open symbols represent
the experimental data measured by the STAR
Collaboration~\cite{40}. The solid and dotted curves are our
results for $K^+$ and $K^-$ fitted by the (a) three-component
Bose-Einstein, (b) three-component Boltzmann, and (c)
single-component Tsallis distributions, respectively.}
\end{figure*}

\begin{figure*}[htb!]
\begin{center}
\includegraphics[width=15.5cm]{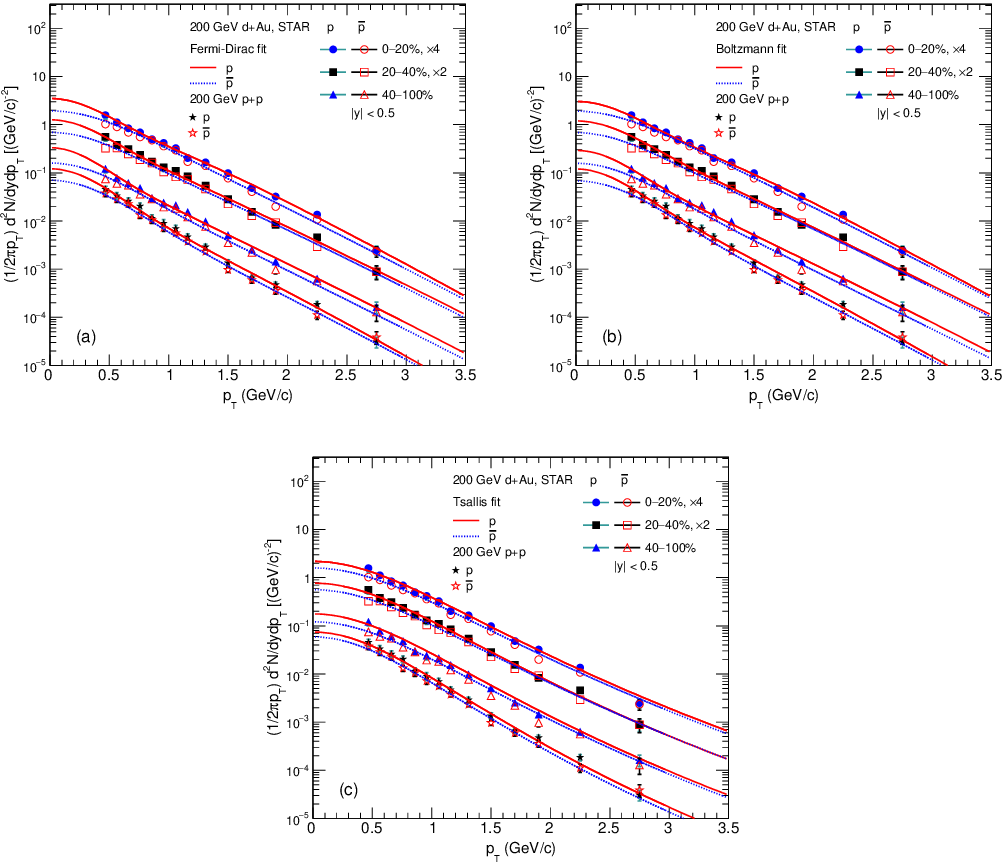}
\end{center}
\justifying\noindent {Figure 3. The invariant yields, $(1/2\pi
p_T)d^2N/dydp_T$, of $p$ and $\bar p$ produced in $|y|<0.5$ in
d+Au collisions with three centrality percentages and in p+p
collisions at $\sqrt{s_{NN}}=200$ GeV. The closed and open symbols
represent the experimental data measured by the STAR
Collaboration~\cite{40}. The solid and dotted curves are our
results for $p$ and $\bar p$ fitted by the (a) three-component
Fermi-Dirac, (b) three-component Boltzmann, and (c)
single-component Tsallis distributions, respectively.}
\end{figure*}

From Figures 1--3 it becomes evident that both three-component
Bose-Einstein (Fermi-Dirac) distributions and Boltzmann
distributions provide a good fit to the transverse momentum
($p_T$) spectra, though a considerable number of parameters were
applied. Due to the application of these three-component standard
distributions, one may obtain more than one value of temperature.
These different values reflect temperature fluctuation from low to
high excitation sources in the framework of multi-source thermal
model~\cite{24,25,26,27,28}. Furthermore, these colorful three- or
multi-component standard distributions can effectively be
approximated by a single-component Tsallis distribution, where the
number of parameters has decreased. The temperature fluctuation
existed in the multi-component standard distributions disappears
in the single-component Tsallis distribution due to the
introduction of index entropy $q$ in the latter.

\subsection{Tendencies of free parameters}

The relationships between effective temperature $T_{eff}$ and
centrality percentage for the production of (a) $\pi^{\pm}$, (b)
$K^{\pm}$, and (c) $p(\bar p)$, obtained from d+Au collisions at
$\sqrt{s_{NN}}=200$ GeV, are illustrated in Figure 4. The
parameter values are derived from Bose-Einstein (Fermi-Dirac,
represented by circles), Boltzmann (squares), and Tsallis
distributions (triangles). It is important to note that the
centrality classes on the horizontal axis are not presented
proportionally. Instead, a schematic diagram has been employed to
facilitate comparison across different values of centrality
percentage. The results depicted in Figure 4 indicate that the
values of $T_{eff}$ derived from the Bose-Einstein, Boltzmann,
Fermi-Dirac, and Tsallis distributions exhibit systematically a
decreasing trend. Meanwhile, these $T_{eff}$ values also show a
decreasing trend with a decrease in collision centrality. In
practical applications, if we consider $T_{\text{Bose-Einstein}}$
and $T_{\text{Fermi-Dirac}}$ as standard baselines, other
temperatures can be evaluated against these benchmarks.

\begin{figure*}[htb!]
\begin{center}
\includegraphics[width=15.5cm]{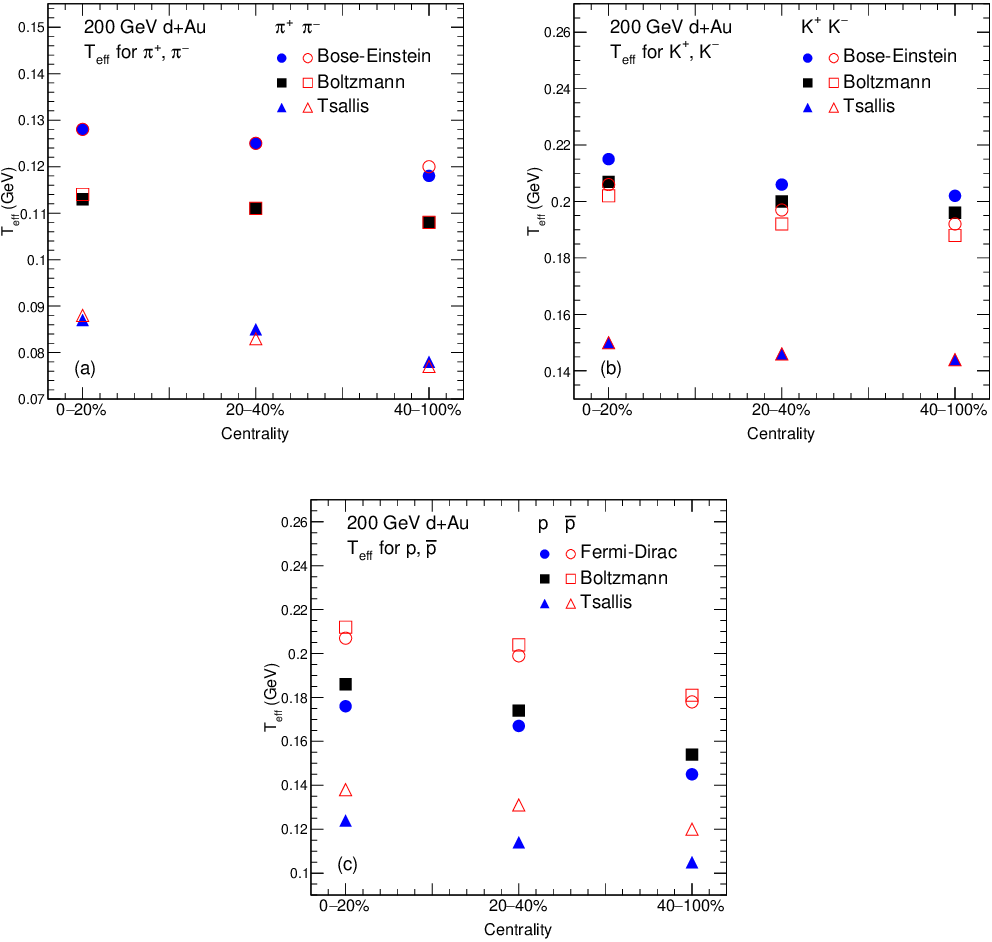}
\end{center}
\justifying\noindent {Figure 4. Relationships between effective
temperature $T_{eff}$ and centrality percentage for productions of
(a) $\pi^{\pm}$, (b) $K^{\pm}$, and (c) $p(\bar p)$ obtained in
d+Au collisions at $\sqrt{s_{NN}}=200$ GeV from the Bose-Einstein
(Fermi-Dirac) (circles), Boltzmann (squares), and Tsallis
distributions (triangles), respectively. Values on the horizontal
axis are not given proportionally. The error bars are invisible
due to their smaller sizes than the symbol one.}
\end{figure*}

\begin{figure*}[htb!]
\begin{center}
\includegraphics[width=9.2cm]{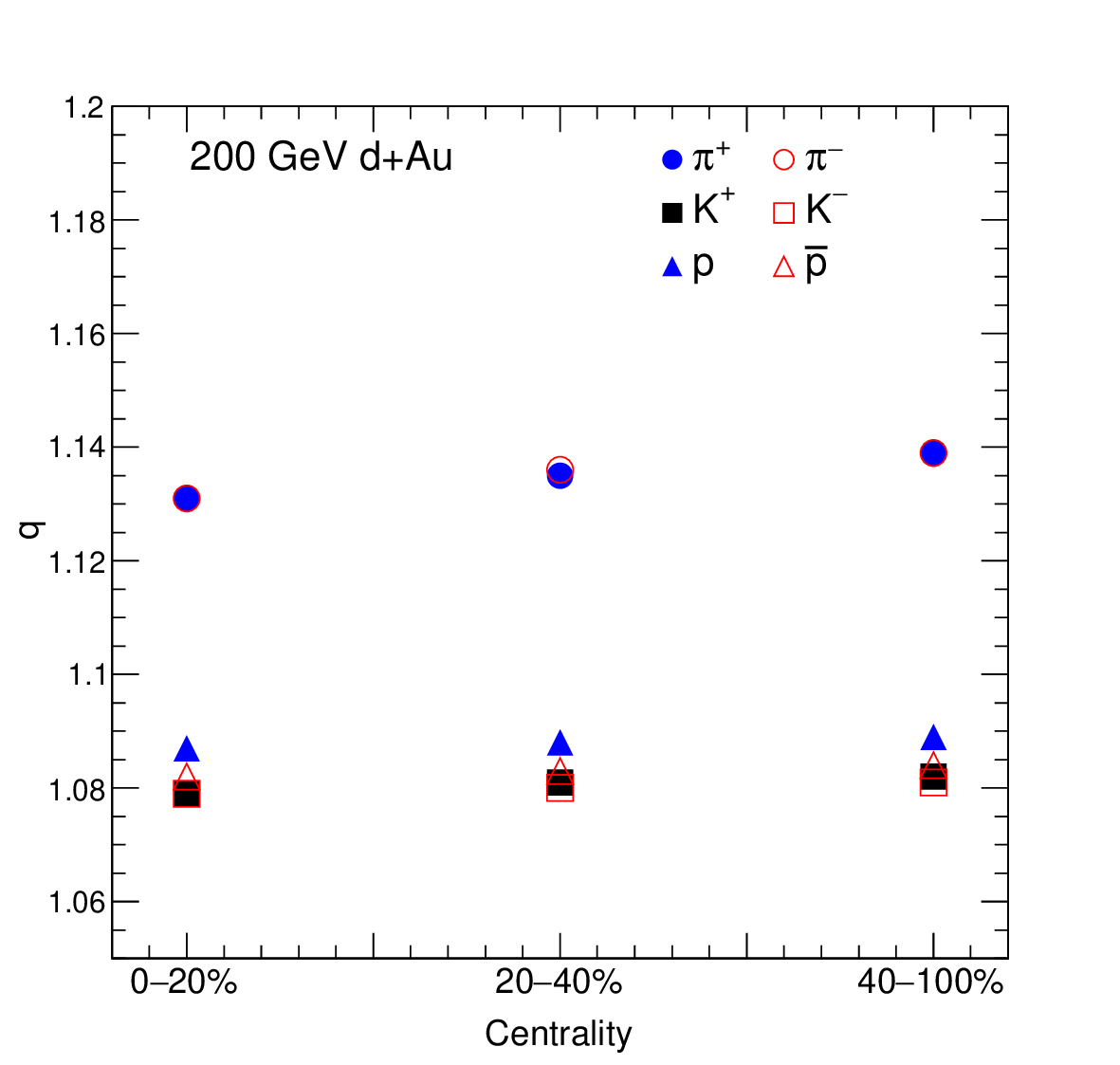}
\end{center}
\justifying\noindent {Figure 5. Relationship between entropy index
$q$ and centrality percentage for the production of $\pi^{\pm}$
(circles), $K^{\pm}$ (squares), and $p(\bar p)$ (triangles)
obtained in d+Au collisions at $\sqrt{s_{NN}}=200$ GeV from the
Tsallis distribution. Values on the horizontal axis are not given
proportionally. The error bars are invisible due to their smaller
sizes than the symbol one.}
\end{figure*}

\begin{figure*}[htb!]
\vspace{-.1cm}
\begin{center}
\includegraphics[width=15.5cm]{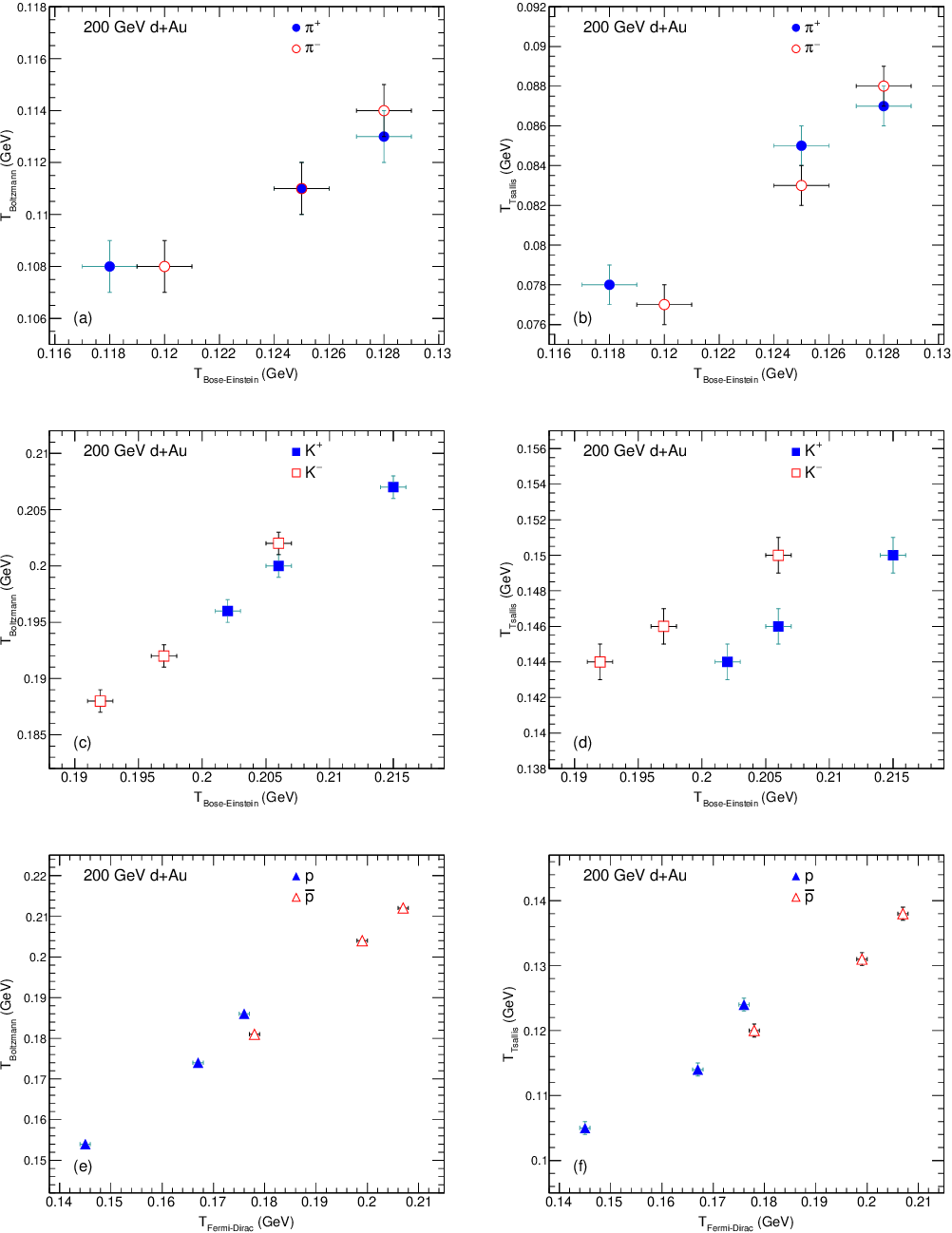}
\end{center}
\justifying\noindent {Figure 6. Relationships between (a)/(b)
$T_{\text{Boltzmann}}/T_{\text{Tsallis}}$ and
$T_{\text{Bose-Einstein}}$ for $\pi^{\pm}$, (c)/(d) $T_{\rm
Boltzmann}$/$T_{\rm Tsallis}$ and $T_{\text{Bose-Einstein}}$ for
$K^{\pm}$, as well as (e)/(f)
$T_{\text{Boltzmann}}/T_{\text{Tsallis}}$ and
$T_{\text{Fermi-Dirac}}$ for $p(\bar p)$ productions in d+Au
collisions at $\sqrt{s_{NN}}=200$ GeV. The closed and open symbols
represent the results for positive and negative hadrons,
respectively.}
\end{figure*}

The relative magnitudes of various effective temperatures suggest
that $T_{\text{Boltzmann}}$ underestimates
$T_{\text{Bose-Einstein}}$ while overestimating
$T_{\text{Fermi-Dirac}}$. This means that
$T_{\text{Bose-Einstein}}>T_{\text{Boltzmann}}$ for particular
bosons and $T_{\text{Boltzmann}}>T_{\text{Fermi-Dirac}}$ for
particular fermions. The reason is the influence of the value of
the last term ($\mp1$) in the denominator of Eq. (1). Likewise, it
appears that the measurement for Tsallis distribution
underestimates both baseline temperatures due to the influence of
entropy index $q$. For particular particles, $T_{\text{Tsallis}}$
is always the minimal one among these effective temperatures.
Furthermore, the effective temperatures shown in Figure 4 clearly
demonstrate an isospin dependence in absolute values for the
production of $K^{\pm}$ due to their differing absorption rates
within hot and dense systems, as well as for the production of
$p(\bar p)$ owing to pre-existing protons present in the colliding
nuclei. These effective temperatures also indicates the multiple
scenarios of kinetic freeze-out with clear mass dependence.
Nevertheless, the trends of these effective temperatures are
consistent.

Figure 5 presents the relationship between entropy index $q$ and
centrality percentage for the production of $\pi^{\pm}$ (circles),
$K^{\pm}$ (squares), and $p(\bar p)$ (triangles). The parameter
values are obtained from d+Au collisions at $\sqrt{s_{NN}}=200$
GeV, but only using Tsallis distribution. Similar to Figure 4, it
should be noted that centrality classes on the horizontal axis are
not displayed proportionally; rather, a schematic representation
has been provided for comparative purposes. Notably, the values of
$q$ derived from the Tsallis distribution approach unity. This
indicates that the collision system is approximately in an
equilibrium state. In comparison to peripheral collisions, $q$ in
central collisions is closer to unity, suggesting that central
collisions are in a more equilibrated state due to more
multi-scatterings in larger participant region. Due to inertia,
more massive particles seem more likely to be in equilibrium.

It should be noted that the $q$ value extracted from $\pi^{\pm}$
spectra is larger than those from $K^{\pm}$ and $p(\bar p)$
spectra for a given centrality. The primary reason lies in the
small pion mass. As the lightest hadron (with a mass of
approximately 140 MeV), the pion undergoes chemical freeze-out
earlier while the system remains in a high-energy density
environment far from equilibrium, thereby exhibiting stronger
non-extensive features in its momentum distribution ($q>1$). In
contrast, the freeze-out times for protons ($\sim938$ MeV) and $K$
mesons ($\sim494$ MeV) occur later, at which point the system
transfers into a phase of relative equilibrium evolution
($q\rightarrow1$). Additionally, due to its small mass and long
mean free path, the pion experiences fewer collisions with other
system components, making its momentum distribution more prone to
deviate from equilibrium (manifested as an increase in the $q$
value). Particles with higher masses tend to thermalize ($q$
closer to 1) through frequent collisions driven by strong
interactions.

Figure 6 illustrates the relationships between (a)/(b)
$T_{\text{Boltzmann}}/T_{\text{Tsallis}}$ and
$T_{\text{Bose-Einstein}}$ for $\pi^{\pm}$, (c)/(d)
$T_{\text{Boltzmann}}/T_{\text{Tsallis}}$ and
$T_{\text{Bose-Einstein}}$ for $K^{\pm}$, as well as (e)/(f)
$T_{\text{Boltzmann}}/T_{\text{Tsallis}}$ and
$T_{\text{Fermi-Dirac}}$ for $p(\bar p)$ productions in d+Au
collisions at $\sqrt{s_{NN}}=200$ GeV. The closed and open symbols
represent results for positive and negative hadrons, respectively.
From Figure 6, it can be observed that there exists an approximate
linear relationship between $T_{\text{Boltzmann}}$ and
$T_{\text{Bose-Einstein}}$ ($T_{\text{Fermi-Dirac}}$), as well as
between $T_{\rm Tsallis}$ and $T_{\text{Bose-Einstein}}$
($T_{\text{Fermi-Dirac}}$). Naturally, there is also an
approximate linear correlation between $T_{\text{Boltzmann}}$ and
$T_{\text{Tsallis}}$~\cite{41}. To draw accurate conclusions
regarding different temperatures with standard baselines, specific
fits related to the hadron spectra are necessary.

In contrast to our previous work~\cite{41}, which examined both
$T_{\rm Boltzmann}$ and $T_{\text{Tsallis}}$ based on data from
large heavy-ion collision systems, this study focuses on effective
temperatures derived from smaller systems such as d+Au and p+p
collisions. By integrating findings from both studies, we may
conclude that not only $T_{\text{Boltzmann}}$ and
$T_{\text{Tsallis}}$, but also $T_{\rm Boltzmann}$ and
$T_{\text{Bose-Einstein}}$ ($T_{\text{Fermi-Dirac}}$), as well as
$T_{\text{Tsallis}}$ and $T_{\text{Bose-Einstein}}$
($T_{\text{Fermi-Dirac}}$), have approximate linear relationships
based on the data with different centrality classes.

\subsection{Further discussion}

To further investigate the relationship between $T_{\rm
Boltzmann}$ and $T_{\text{Bose-Einstein}}$
($T_{\text{Fermi-Dirac}}$), along with $T_{\rm Tsallis}$ and
$T_{\text{Bose-Einstein}}$ ($T_{\text{Fermi-Dirac}}$), we utilize
the parameters $q$ and $V$ derived from 0--20\% d+Au collisions at
$\sqrt{s_{NN}}=200$ GeV. The results of the Tsallis distributions
for production of $\pi^+$, $K^+$, and $p$ are obtained over a
range of $T_{\text{Tsallis}}$ from 0.006 to 0.26 GeV in increasing
increments of 0.002, 0.01, or 0.02 GeV, which are subsequently
fitted using three-component Bose-Einstein (Fermi-Dirac) and
Boltzmann distributions. Selected examples of these Tsallis
distributions at specific values of $T_{\text{Tsallis}}=0.10$,
0.14, 0.18, and 0.22 GeV, along with their corresponding fitting
results, are illustrated in Figures 7(a)--7(d). The meanings of
various curves are indicated within each panel. It is evident that
the Tsallis distribution can be effectively fitted by
three-component Bose-Einstein (Fermi-Dirac) and Boltzmann
distributions across a given range of $p_T$, although some
deviations occur in both low- and high-$p_T$ regions. Similar
results for $\pi^-$, $K^-$, and $\bar p$ have been observed as
shown in Figure 7. However, they are not presented here to avoid
redundancy.

\begin{figure*}[htb!]
\vspace{-.1cm}
\begin{center}
\includegraphics[width=15.5cm]{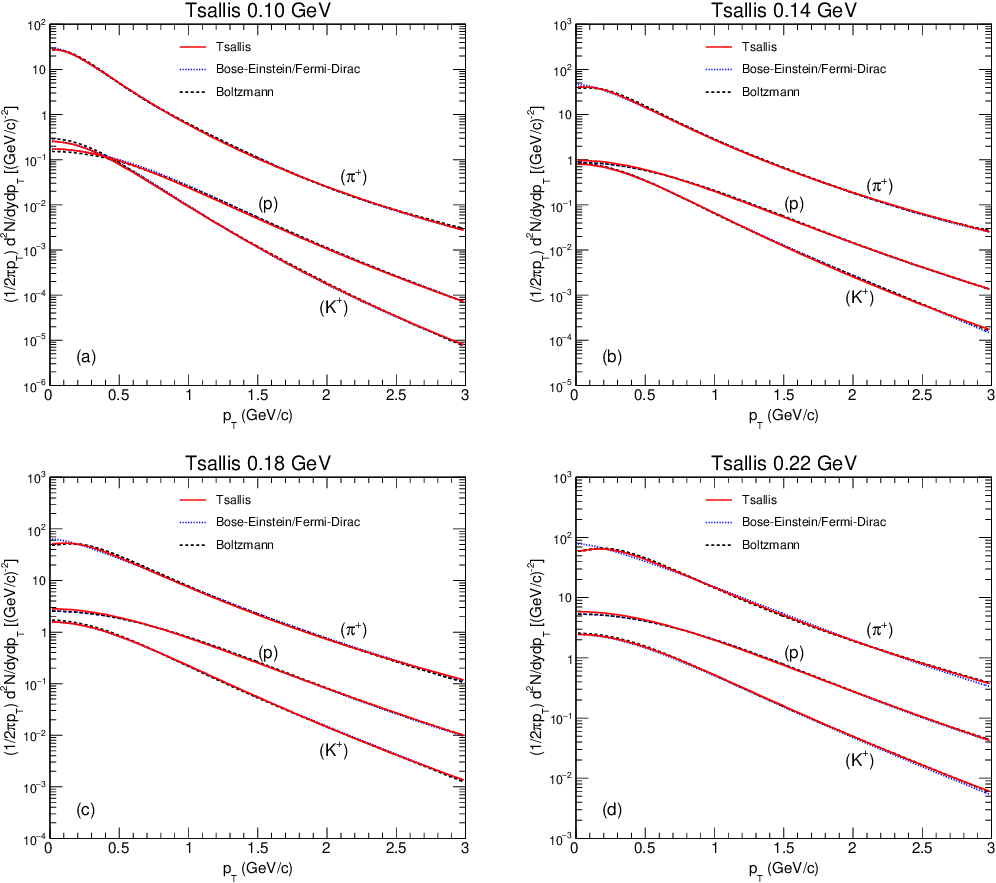}
\end{center}
\justifying\noindent {Figure 7. The selected Tsallis distributions
with $T_{\text{Tsallis}}=0.10$ (a), 0.14 (b), 0.18 (c), and 0.22
GeV (d) and the related fittings by three-component Bose-Einstein
(for $\pi^+$ and $K^+$), Fermi-Dirac (for $p$), and Boltzmann
distributions. The values of $q$ and $V$ from 0--20\% d+Au
collisions at $\sqrt{s_{NN}}=200$ GeV are used in Tsallis
distributions with different $T_{\text{Tsallis}}$. The meanings
referred to various curves are marked in the panels.}
\end{figure*}

\begin{figure*}[htb!]
\vspace{-.1cm}
\begin{center}
\includegraphics[width=15.5cm]{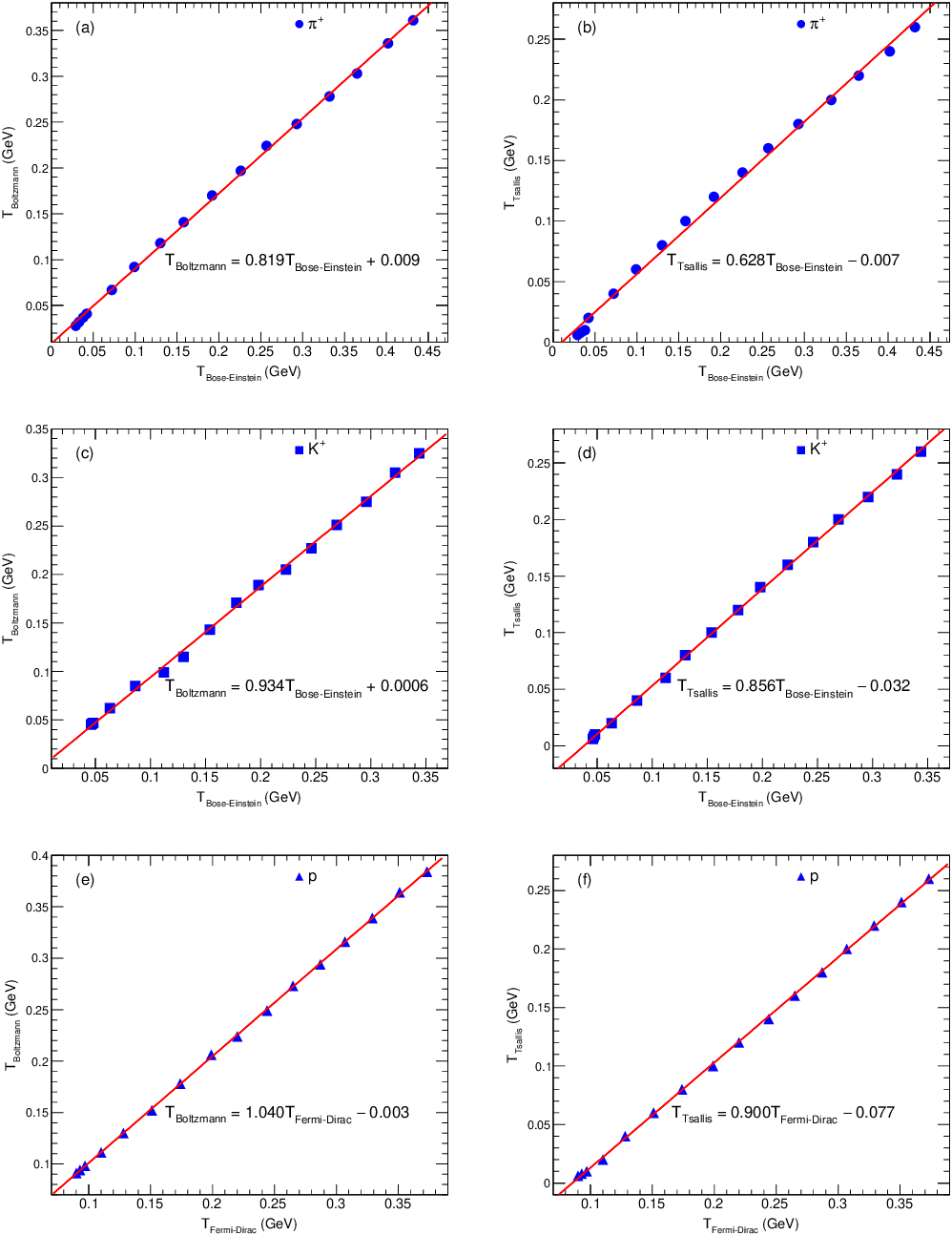}
\end{center}
\justifying\noindent {Figure 8. Relationships between (a)/(b)
$T_{\text{Boltzmann}}/T_{\text{Tsallis}}$ and
$T_{\text{Bose-Einstein}}$ for $\pi^+$, (c)/(d) $T_{\rm
Boltzmann}$/$T_{\rm Tsallis}$ and $T_{\text{Bose-Einstein}}$ for
$K^-$, as well as (e)/(f)
$T_{\text{Boltzmann}}/T_{\text{Tsallis}}$ and
$T_{\text{Fermi-Dirac}}$ for $p$ productions. The symbols
represent the selected 16 $T_{\rm Tsallis}$ which equals to 0.006,
0.008, 0.01, 0.02, 0.04, 0.06, ..., 0.26 GeV, for which $q$ and
$V$ are taken from 0--20\% d+Au collisions at $\sqrt{s_{NN}}=200$
GeV. The straight lines are the fitting results with the linear
equations shown in the panels.}
\end{figure*}

The relationships depicted in panels (a)/(b), showing the
correlation between $T_{\text{Boltzmann}}/T_{\text{Tsallis}}$ and
$T_{\text{Bose-Einstein}}$ for $\pi^+$; panels (c)/(d),
illustrating the same correlation for $K^+$; as well as panels
(e)/(f), representing the correlation between
$T_{\text{Boltzmann}}/T_{\text{Tsallis}}$ and
$T_{\text{Fermi-Dirac}}$ for $p$ productions, concerning various
temperatures based on $q$ and $V$ values obtained from 0--20\%
d+Au collisions at $\sqrt{s_{NN}}=200$ GeV---are provided in
Figures 8(a)--8(f). In these figures, symbols denote various
effective temperatures $T_{eff}$ utilized within the
three-component Bose-Einstein (Fermi-Dirac) and Boltzmann
distributions alongside those used in the Tsallis distribution
itself where applicable; the straight lines are the fitting
results with the linear equations shown in the panels. Notably, a
very good linear relationship for given correlation and particle
spectra can be observed across all six panels presented, though
three-component standard distributions are employed to fit the
Tsallis distribution.

The Bose-Einstein and Fermi-Dirac distributions are widely
acknowledged as standard baselines in both classical and modern
physics, providing an accurate representation of effective
temperature. In contrast, the Boltzmann distribution serves as a
reasonable approximation; however, it tends to slightly
underestimate (or overestimate) the effective temperature derived
from the Bose-Einstein (or Fermi-Dirac) distribution. In addition
to these standard distributions, the Tsallis distribution offers
an alternative framework. However, it significantly underestimates
the effective temperature obtained from three-component
Bose-Einstein or Fermi-Dirac distributions due to its
incorporation of an entropy index. While other distributions such
as the q-dual distribution~\cite{20} and Erlang
distribution~\cite{24,25,28} can be utilized to extract effective
temperatures~\cite{42,43}, establishing a connection between these
extracted effective temperatures and those derived from standard
distributions is essential for meaningful comparisons.

Our recent findings~\cite{43} indicate an approximate linear
relationship or positive correlation---rather than an exact linear
correlation---between $T_{\text{Boltzmann}}$ and
$T_{\text{Bose-Einstein}}$ (or $T_{\text{Fermi-Dirac}}$), as well
as between $T_{\text{Tsallis}}$ and $T_{\text{Bose-Einstein}}$ (or
$T_{\text{Fermi-Dirac}}$) from a comprehensive analysis on the
spectra of light hadrons produced in gold-gold (Au+Au) collisions
with different centralities over center-of-mass energy range from
7.7 to 200 GeV. The present work shows better linear correlations
based on given particle spectra from 0--20\% d+Au collisions at
200 GeV. This outcome is in line with our earlier
expectation~\cite{41} of an approximate precise linear
relationship among Tsallis and Boltzmann temperatures, from Au+Au
collisions at 130 and 200 GeV, as well as lead-lead (Pb+Pb)
collisions at 2.76 TeV with various centralities. Regardless of
which effective temperature is employed, it is appropriate to
adopt the effective temperature derived from standard
distributions as a uniform baseline. Numerically, at the same
$\sqrt{s_{NN}}$, the effective temperature extracted from central
nucleus-nucleus collisions exceeds that from peripheral
collisions, whereas the effective temperature derived from p+p
collisions is closer to and slightly lower than that from
peripheral collisions.

Before summarizing and concluding, we want to point out that, when
applied to the same dataset, the results obtained using this
multi-component ideal gas method are more readily connected to
traditional thermodynamics compared to those derived from
alternative or more sophisticated models, although a one-to-one
correspondence can be established between the two approaches.
Moreover, despite high-energy collision processes displaying
properties of a small-volume QGP evolution, the behavior of the
final-state particles exhibits characteristics of the large-volume
gases, validating the rationality of applying the multi-component
ideal gas method to study final-state particle behavior. We have
analyzed the transverse momentum spectra of various particles
across collision processes with different energies and systems,
achieving reasonable fitting results~\cite{41,43}, which further
corroborates the validity of the method employed. During the
fitting process, we used a combination of least squares method and
leave-one-out cross validation method to determine the optimal
parameters.

It is essential to highlight the connection between effective
temperature $T_{eff}$ measurements and the QGP phase transition.
While QGP formation occurs on $\sim1$ fm/$c$ time scales in
high-energy collisions, hydrodynamic evolution preserves memory of
initial conditions via flow observables. The $T_{eff}$ extracted
from final-state particle spectra reflects the collective
expansion dynamics initiated during the QGP phase. The observed
non-monotonic dependence of $T_{eff}$ on collision
energy/centrality provides indirect evidence of the phase
transition and potential critical endpoint signatures. Of course,
studying phase transitions and critical endpoints cannot be
decoupled from temperatures. However, temperatures are
model-dependent and necessitate a standardized baseline for
meaningful comparisons. Regrettably, no specific criteria or
constraints currently exist within the community to ensure
consistent comparisons across models. We propose adopting
Bose-Einstein and Fermi-Dirac temperatures, extracted from
multi-component distributions and widely used in physics for an
extended period, as the standard baseline.

We would like to emphasize that this study overcomes the
limitations of traditional thermal models---typically confined to
large systems---by systematically comparing effective temperatures
derived from Bose-Einstein (Fermi-Dirac), Boltzmann, and Tsallis
distributions for light hadron transverse momentum spectra in
small collision systems at top RHIC energy. This research
transcends conventional thermal models, offering an important lens
to probe the interrelation between non-equilibrium and local
equilibrium phenomena in Quantum Chromodynamics (QCD) systems.
Indeed, the analysis reveals the coexistence of pronounced
non-equilibrium effects for single-event in small systems and
local thermal equilibrium for lots of events. While quantum
statistics (Bose-Einstein/Fermi-Dirac) and Tsallis non-extensive
statistics diverge in describing spatiotemporally constrained
systems, their applications highlight the persistence of partial
thermalization. This dual behavior challenges the universality of
equilibrium assumptions and underscores the need for hybrid
statistical frameworks to decode the QCD medium's evolution.

The interplay between Tsallis entropy index $q$ and effective
temperatures $T$ establishes a quantitative link between
microscopic quantum fluctuations (e.g., fractal phase-space
structures) and macroscopic thermodynamic observables. Such
correlations suggest fractal thermalization as a potential feature
of QCD matter under extreme conditions. Furthermore, anchoring
effective temperatures to Bose-Einstein (Fermi-Dirac) baselines
provides a foundational reference for cross-model comparisons,
enabling unified interpretations of temperature fluctuations
across statistical paradigms. By demarcating the dominance of
quantum statistics at low $p_T$ and Tsallis non-extensivity at
high $p_T$, this work guides the design of precision momentum
spectrum measurements at RHIC, Large Hadron Collider (LHC), and
future Electron-Ion Collider (EIC) facilities. These insights can
not only provide reference for optimizing detector design, but
also improve hydrodynamic and transport models to more accurately
capture non-equilibrium phase transitions. Ultimately, the
integration of multi-model perspectives can reshape theoretical
benchmarks for thermalization criteria, and advance the
exploration of exotic quantum matter in high-energy collisions and
early-universe.

\section{Summary and conclusions}

The transverse momentum spectra of $\pi^{\pm}$, $K^{\pm}$, and
$p(\bar p)$ produced in d+Au and p+p collisions at
$\sqrt{s_{NN}}=200$ GeV, measured by the STAR Collaboration, are
analyzed using three-component Bose-Einstein (Fermi-Dirac),
three-component Boltzmann, and single-component Tsallis
distributions. The effective temperatures $T_{eff}$ and entropy
index $q$ derived from the spectra of different particles reveal a
multi-scenario approach to kinetic freeze-out. Due to the
significantly high yields of $\pi^{\pm}$, the average parameters
weighted across different particles closely resemble those
obtained from the spectra of $\pi^{\pm}$.

Our findings indicate that the values of $T_{eff}$ derived from
three-component Bose-Einstein, Boltzmann, and Fermi-Dirac
distributions, as well as from single-component Tsallis
distribution, exhibit systematically a decreasing trend.
Meanwhile, these $T_{eff}$ values also show a decreasing trend
with a decrease in collision centrality. In practical
applications, if we consider $T_{\text{Bose-Einstein}}$ and
$T_{\text{Fermi-Dirac}}$ as standard baselines for comparison with
other temperature measurements, it is observed that
$T_{\text{Boltzmann}}$ underestimates $T_{\text{Bose-Einstein}}$
while overestimating $T_{\text{Fermi-Dirac}}$, whereas
$T_{\text{Tsallis}}$ underestimates both temperatures.
Furthermore, a nearly perfect linear relationship exists between
$T_{\text{Boltzmann}}$ and $T_{\text{Bose-Einstein}}$
($T_{\text{Fermi-Dirac}}$), as well as between $T_{\rm Tsallis}$
and $T_{\text{Bose-Einstein}}$ ($T_{\text{Fermi-Dirac}}$), based
on given particle spectra.

As one moves from central to peripheral d+Au collisions, there is
a slight decrease in the values of $T_{eff}$ obtained from various
distributions for emissions across different particles.
Concurrently, values of $q$ extracted from the Tsallis
distribution show a gradual increase for emissions involving
different particles. Notably, parameters obtained from p+p
collisions exhibit similarities to those observed in peripheral
d+Au collisions. This similarity arises from nearly identical
energy depositions, comparable initial-state geometries, and
minimized hadronic re-scattering effects in both p+p and d+Au
collisions systems.
\\
\\
{\bf Acknowledgments}

The work of Shanxi group was supported by National Natural Science
Foundation of China under Grant No. 12147215, Fundamental Research
Program of Shanxi Province under Grant No. 202303021221071, Shanxi
Scholarship Council of China and the Fund for Shanxi ``1331
Project" Key Subjects Construction. The work of P.-P.Y. was
supported by Fundamental Research Program of Shanxi Province under
Grant No. 202203021222308, Doctoral Scientific Research
Foundations of Shanxi Province and Xinzhou Normal University, and
Academic Leading Specialist Project of Xinzhou Normal University
under Grant Nos. 2024RC10 and 2024RC10B. The work of K.K.O. was
supported by the Agency of Innovative Development under the
Ministry of Higher Education, Science and Innovations of the
Republic of Uzbekistan within the fundamental project No.
F3-20200929146 on analysis of open data on heavy-ion collisions at
RHIC and LHC.
\\
\\
{\bf Data availability}

The data used to support the findings of this study are included
within the article and are cited at relevant places within the
text as references.
\\
\\
{\bf Declarations}
\\
\\
{\bf Ethical approval}

The authors declare that they are in compliance with ethical
standards regarding the content of this paper.
\\
\\
{\bf Conflict of interest}

The authors declare that there are no conflict of interest
regarding the publication of this paper. The funding agencies have
no role in the design of the study; in the collection, analysis,
or interpretation of the data; in the writing of the manuscript;
or in the decision to publish the results.
\\

\end{document}